\documentclass[12pt]{article}
\usepackage[utf8]{inputenc}
\usepackage[margin=0.82in]{geometry}
\usepackage{colortbl}
\usepackage{multicol}
\usepackage{multirow}
\usepackage{hhline}
\usepackage{authblk}

\usepackage[T1]{fontenc}    
\usepackage{hyperref}       
\usepackage{url}            
\usepackage{booktabs}       
\usepackage{amsfonts}       
\usepackage{nicefrac}       
\usepackage{microtype}      
\usepackage{wrapfig}
\usepackage{lipsum}

\usepackage{url,amsmath,amsfonts,amssymb,amsthm,graphics,graphicx,color}

\usepackage{verbatim}

\usepackage{longtable}

\usepackage{xr}

\title{\bf SARS-CoV-2 orthologs of pathogenesis-involved small viral RNAs of SARS-CoV}

\author[]{Ali Ebrahimpour Boroojeny} 
\author[]{Hamidreza Chitsaz}
\affil[]{Department of Computer Science, Colorado State University}

\date{\texttt{http://chitsazlab.org}\\ \texttt{chitsaz@chitsazlab.org}}

\begin{document}

\maketitle

\begin{abstract}
{\bf Background} The COVID-19 pandemic clock is ticking and the survival of many of mankind's modern institutions and or survival of many individuals is at stake. There is a need for treatments to significantly reduce the morbidity and mortality of COVID-19.

Hence, we delved deep into the SARS-CoV-2 genome, which is the virus that has caused COVID-19. SARS-CoV-2 is from the same family as SARS-CoV in which three small viral RNAs (svRNA) were recently identified \cite{morales2017sars}; those svRNAs play a significant role in the virus pathogenesis in mice. 

{\bf Contribution} In this paper, we report potential orthologs of those three svRNAs in the SARS-CoV-2 genome. Instead of off-the-shelf search and alignment algorithms, which failed to discover the orthologs, we used a special alignment scoring that does not penalize C/T and A/G mismatches as much as the other mutations. RNA bases C and U both can bind to G; similarly, A and G both can bind to U, hence, our scoring. We also validate this hypothesis using a novel, independent computational experiment.

To validate our results, we confirmed the discovered orthologs are fully conserved in all the tested publicly available genomes of various strains of SARS-CoV-2; the loci at which the SARS-CoV-2 orthologs occur are close to the loci at which SARS-CoV svRNAs occur. We also report potential targets for these svRNAs. We hypothesize that the discovered orthologs play a role in pathogenesis of SARS-CoV-2, and therefore, antagomir-mediated inhibition of these SARS-CoV-2 svRNAs inhibits COVID-19. 

\end{abstract}

\section{Introduction}

The world is now struggling with a pandemic known as COVID-19 which is caused by a novel coronavirus that was first identified in December 2019 in a local sea food market in Wuhan, China \cite{luoutbreak}. Due to the similarity of its genomic sequence to that of Severe Acute Respiratory Syndrome (SARS-CoV), which is a member of the subgenus of Sarbecovirus, the aforementioned novel coronavirus was named SARS-CoV-2. Phylogenetic studies have found a bat origin for this virus \cite{lu2020genomic,wan2020receptor}. As of April $29^{\mbox{th}}, 2020$, this virus has infected more than 45,360,000 people in 190 countries, caused more than 1,185,000 cases of death, and has become a global health concern leading to massive lock downs and quarantine all around the world.  

Since the emergence of SARS-CoV in China in 2002, which infected around 8,000 people world-wide, multiple research efforts have tried to understand that virus and to suggest potential treatments. Despite the fact that no vaccines or antivirals have been approved to date for any of coronaviruses, improvements on reducing the severity of the disease and mortality rate have been reported. Because of the similarity of the recent fast-spreading coronavirus SARS-CoV-2, which shares more than $79\%$ of its genomic sequence with SARS-CoV, one plausible way to understand how it works and suggest possible treatments would be to port what has been found for SARS-CoV previously to SARS-CoV-2.

Non-coding RNAs (ncRNAs) are, as the name suggests, RNAs that do not translate to proteins. Although it is likely the case that some of them do not play a major role in the cell \cite{brosius2005waste,palazzo2015non}, some have crucial functions, such as transfer RNAs (tRNAs), ribosomal RNAs (rRNAs), micro RNAs (miRNAs), etc. Some of the ncRNAs, such as miRNAs, play a role in post-transcriptional regulation of gene expression. They, through a procedure called gene silencing, bind with the complementary parts of the target RNAs, and prevent the translation of those RNAs through cleavage of their strand, shortening their poly-A tail, or downgrading the efficiency of their translation by making some nucleotides unavailable to the ribosomes \cite{fabian2010regulation, bartel2009micrornas}.   

First viral ncRNA was identified by Reich \textit{et al.} \cite{reich1966rna}. Since then a plethora of viral-associated ncRNAs have been identified and this has been accelerated by the advances in technology \cite{tycowski2015viral}. Especially, deep sequencing has facilitated the detection of small virus-associated RNAs \cite{parameswaran2010six, perez2010influenza}. Some of these ncRNAs are known to be responsible in counteracting the antiviral defense mechinism that are present in the host cells, mostly through inhibition of protein kinase R (PKR) \cite{steitz2011noncoding}. Therefore, they aid in the life cycle of the virus \cite{banerjee1987transcription, cullen2009viral}, such as svRNAs in influenza A virus that are involved in the mechanisms this virus uses for switching between transcription and replication \cite{perez2010influenza}.

It had been well-known that nuclear and DNA viruses encode miRNAs \cite{cullen2011viruses} that play a role in persistence \cite{tenoever2013rna} of the virus as well as changing the transcriptopme in the host cell \cite{bartel2004micrornas}. Using the deep sequencing technologies, it had been revealed that cytoplasmic RNA viruses also express ncRNAs \cite{parameswaran2010six, perez2010influenza, morales2017sars} and most of them induce various cytoplasmic pathways to express their ncRNAs \cite{perez2010influenza}. Flaviviruses can be mentioned as examples of cytoplasmic RNA viruses, which are very sensitive to interferons and have evolved a variety of mechanisms to avoid their action \cite{diamond2009mechanisms}. It has been shown that ncRNAs in flaviviral RNA bounds to genes responsible for regulation of antiviral state of the host cell and affects the interferon response agains the virus \cite{bidet2014g3bp1}.

A recent research has reported three small viral RNAs (svRNAs) that are derived from the genomic regions of SARS-CoV \cite{morales2017sars}. Morales \textit{et al.} have have shown the presence of these positive sense svRNAs, which are ``mapped to nsp3 at the $5^\prime$ end of the Replicase gene and the N gene (svRNA-N) at the $3^\prime$ end of the genomic RNA (gRNA)'' \cite{morales2017sars}, by using specific small RNA RT-qPCR assays. Their experiments on a mouse model of the infection \cite{roberts2007mouse, dediego2007severe} show that these svRNAs contribute to SARS-CoV pathogenesis, and also suggest a potential antiviral treatment using antagomir-mediated inhibition of these svRNAs. 

Small non-coding RNAs that play an improtant roll have been shown to be highly conserved among genuses and families. Given that SARS-CoV-2 is in the the same subgenus as SARS-CoV and their genomic sequence has more than $79\%$ similarity, if orthologs of the svRNAs found in SARS-CoV be present and found in SARS-CoV-2, then (antagomir-mediated) inhibition of those svRNA orthologs is expected to lead to reduction of SARS-CoV-2 titers and help decrease severity of COVID-19 in the majority of patients, as was shown to be the case for SARS-CoV \cite{morales2017sars}. What is needed is the sequence of the three aforementioned svRNAs in SARS-CoV-2 genome, from which the corresponding antagomirs are simply designed through base pair complementarity.

To target SARS-CoV-2 svRNAs, we first characterize the sequence of three svRNAs in SARS-CoV-2. We achieve that goal by aligning the sequences of SARS-CoV svRNAs to the SARS-CoV-2 genome sequences. After investigating many of the available sequenced genomes of SARS-CoV-2 that have been reported in various locations around the globe, we discovered the presence of three svRNAs that were highly conserved orthologs of the svRNAs that played a role in pathogenesis of SARS-CoV. This \emph{in silico} discovery still needs to be confirmed using \emph{in vitro} and \emph{in vivo} experiments, but our findings reported in the following sections suggest strong likelihood of our hypothesis.

\section{Methods}\label{sec:methods}

In order to find the svRNAs in SARS-Cov-2 that are orthologs of those in SARS-CoV, we selected all of the complete genomic sequences of SARS-CoV-2, that were made available on the NCBI portal as of March $27^{th}$ (173 complete sequences) as well as 27 more randomly chosen ones among the more recent uploads on the NCBI portal. These genomic sequences are from different states and countries, including but not limited to: New York, Washington, California, Illinois, Utah, China, Japan, South Korea, Italy, India, Brazil, Germany, Australia, Turkey, and Greece.

We used a variant of the algorithm introduced by Smith and Waterman, which is known in the community as the \textit{fit alignment}\cite{Smith81}, as the core of investigation for the svRNAs of interest. The algorithm devised by Smith and Waterman is itself a variant of the \textit{global alignment} algorithm known as Needleman-Wunsch algorithm \cite{needleman1970general}, and can be used to find the regions of two genomic sequences that are similar. Fit alignment is a variant of this algorithm that searches a reference genome for a subsection that is highly similar to another shorter sequence. 

We used our own in-house alignment tool, which is freely availabe \footnote{https://github.com/Ali-E/WeightedAligner}, that implements a faster variant of local and fit alignment algorithms, based on the idea of limiting the search to considering only a subset of the alignment search space that represents high similarity and pruning the regions that fall below the desired threshold. Also, when aligning two sequences, there might be multiple alignments with the same score as the highest score. Therefore, our tool keeps track of all the alignments with a score equal to the highest score. For this specific problem, we also added the feature of choosing the best alignment that exists in all the other reference sequences. However, we should note that, for all the sequences considered (the ones in table~\ref{tab:match}) the loci reported for the exact match to our suggested svRNAs have also the highest score of alignment with the svRNAs reported by Morales \textit{et al.} in their respective sequence.

Another distinctive feature of our alignment tool is the ability to penalize A/G and C/T mismatches less than the other mismatches. The intuition behind this feature is that bases A and G both can bind to U, so by mutating A to G, GU bonds can replace AU bonds. The same idea holds for C and T bases which can bind to G bases. To further test this hypothesis we used 1000 randomly chosen pairs of interacting RNAs from the RISE database. After extracting the reported binding sites, we made 3 other sets of sequence-pairs by randomly mutating one of the A bases of each of the sequences to C, G, and T (the position of the mutating A is the same); we repeat this random mutation 5 separate times (each time a potentially different A base in a sequence is chosen for mutation) to increase the number of samples in each set and have a more unbiased generalization. After making this 3 sets of mutated versions of the original data, we compute the energy of each of these pairs using piRNA~\cite{Chitsaz09} which is a tool that computes free energy of a pair of RNAs by considering both the enthalpy and entropy of the interaction RNA structures. Finally, we compare the energy changes to see which mutations are less detrimental than the others.

\section{Results}

Figures 1, 2, and 3 show the sequences that we have found for three svRNAs in SARS-CoV-2 that are orthologs of the three aforementioned svRNAs in SARS-CoV. To further test our hypothesis, we searched for all these svRNAs in 200 different complete reference sequences of the virus. Our three svRNAs are wholly present, without any mutation, in all the reference sequences. Table \ref{tab:match} shows the NCBI ID of each of these sequences, as well as the string loci of each of the svRNAs. As you can see 199 out of 200 tested sequences are present in the table and contain the exact match of the proposed RNAs. The missing entry of the table is LR757997 which showed the presence of the third svRNA at loci 28604 but does not contain the other two because there is a gap in the sequence from loci 3001 to 3235 (filled with Ns), and this is the region where the first two svRNAs reside in according to the loci values in the table for these two columns.

We identify a non-detrimental mismatch by $:$ in Figures 1, 2, and 3. RNA-RNA binding energies is mainly governed by Watson-Crick base pairing, namely A-U, G-U, and C-G. Particularly, U can pair with both A and G. Hence, an A vs. G mismatch (substitution) in an ortholog RNA is non-detrimental for the binding to a target RNA. Similarly, G can pair with both U and C. Hence, an U/T vs. C mismatch (substitution) in an ortholog RNA is non-detrimental for the binding to a target RNA.

\begin{figure}[h!]
\begin{center}
\includegraphics[width=0.45\textwidth]{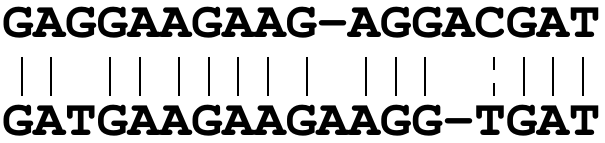}
\end{center}
\caption {Alignment of the first svRNA in SARS-CoV and its identified ortholog in SARS-CoV-2. Top: nsp3.1 svRNA sequence GAGGAAGAAGAGGACGAT in SARS-CoV according to \cite{morales2017sars}. Bottom: Ortholog of nsp3.1 svRNA sequence GATGAAGAAGAAGGTGAT in SARS-CoV-2. $|$ represents a match, $-$ represents a gap (indel), empty represents a mismatch, and $:$ represents a non-detrimental mismatch. \label{fig:svRNA1}} 
\end{figure}

\begin{figure}[h!]
\begin{center}
\includegraphics[width=0.55\textwidth]{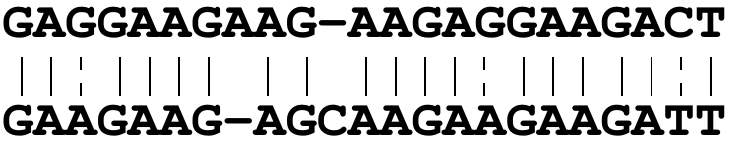}
\end{center}
\caption {Alignment of the second svRNA in SARS-CoV and its identified ortholog in SARS-CoV-2. Top: nsp3.2 svRNA sequence GAGGAAGAAGAAGAGGAAGACT in SARS-CoV according to \cite{morales2017sars}. Bottom: Ortholog of nsp3.2 svRNA sequence GAAGAAGAGCAAGAAGAAGATT in SARS-CoV-2. $|$ represents a match, $-$ represents a gap (indel), empty represents a mismatch, and $:$ represents a non-detrimental mismatch. \label{fig:svRNA2}} 
\end{figure}

\begin{figure}[h!]
\begin{center}
\includegraphics[width=0.55\textwidth]{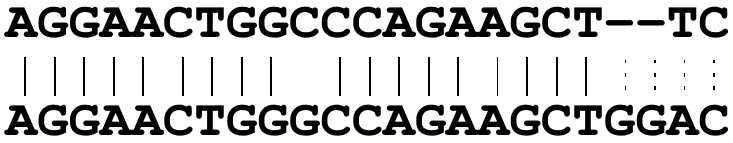}
\end{center}
\caption {Alignment of the third svRNA in SARS-CoV and its identified ortholog in SARS-CoV-2. Top: N svRNA sequence AGGAACTGGCCCAGAAGCTTC in SARS-CoV according to \cite{morales2017sars}. Bottom: Ortholog of N svRNA sequence AGGAACTGGGCCAGAAGCTGGAC in SARS-CoV-2. $|$ represents a match, $-$ represents a gap (indel), empty represents a mismatch, and dotted line represents a potential omission. It is possible that some or all of the last four (4) nucleotides GGAC of the bottom sequence are dropped (omitted). \label{fig:svRNA3}} 
\end{figure}

The table below shows the results of the experiment designed to evaluate our hypothesis about non-detrimental mutations. The tables show the amount of decrease in the free energy which results in a more stable structure. Therefore the higher the decrease value gets, the more energetically favorable the resulting structure becomes after the mutation. The results in the first table show that A/G mutations are the least detrimental ones and they even lead to a more favorable interaction structure ensemble than the original ones. This can be explained by the analysis reported by Boroojeny et al.~\cite{Ebrahimpour20d} in which authors concluded that CG bonds are stronger than GU bonds, and they are both stronger than AU bonds. Therefore when A is mutated to G, the potentially existing AU interaction bond can be at least replaced by a GU bond which is stronger; that G even finds opportunity to form a GC bond with a free C base which would theoretically be even stronger. 

The second table reports the results for C mutations. Based on the table, we can see that C/T mutations are not as detrimental as C/A mutations, but still seem to be more detrimental than C/G mutations. However, the scatter plot in Figure~\ref{fig:CT} shows that for the more energetically favorable interaction structures (the ones toward the right side of the plot), the C/T mutations seem to be less detrimental than C/G mutations, which results in the intersection of the fitted lines around -8 kcal/mol. Based on these results, we penalize C/T mutations slightly more than A/G mutations when finding the best alignments.

\begin{center}
\begin{tabular}{ c | c | c | c }
 Mutation & A/G & A/C & A/T\\ 
 \hline
 Decrease of the free energy & 2.27 & 0.44 & 0.50\\ 
 
\end{tabular}
\end{center}

\begin{figure}[h!]
\begin{center}
\includegraphics[width=0.55\textwidth]{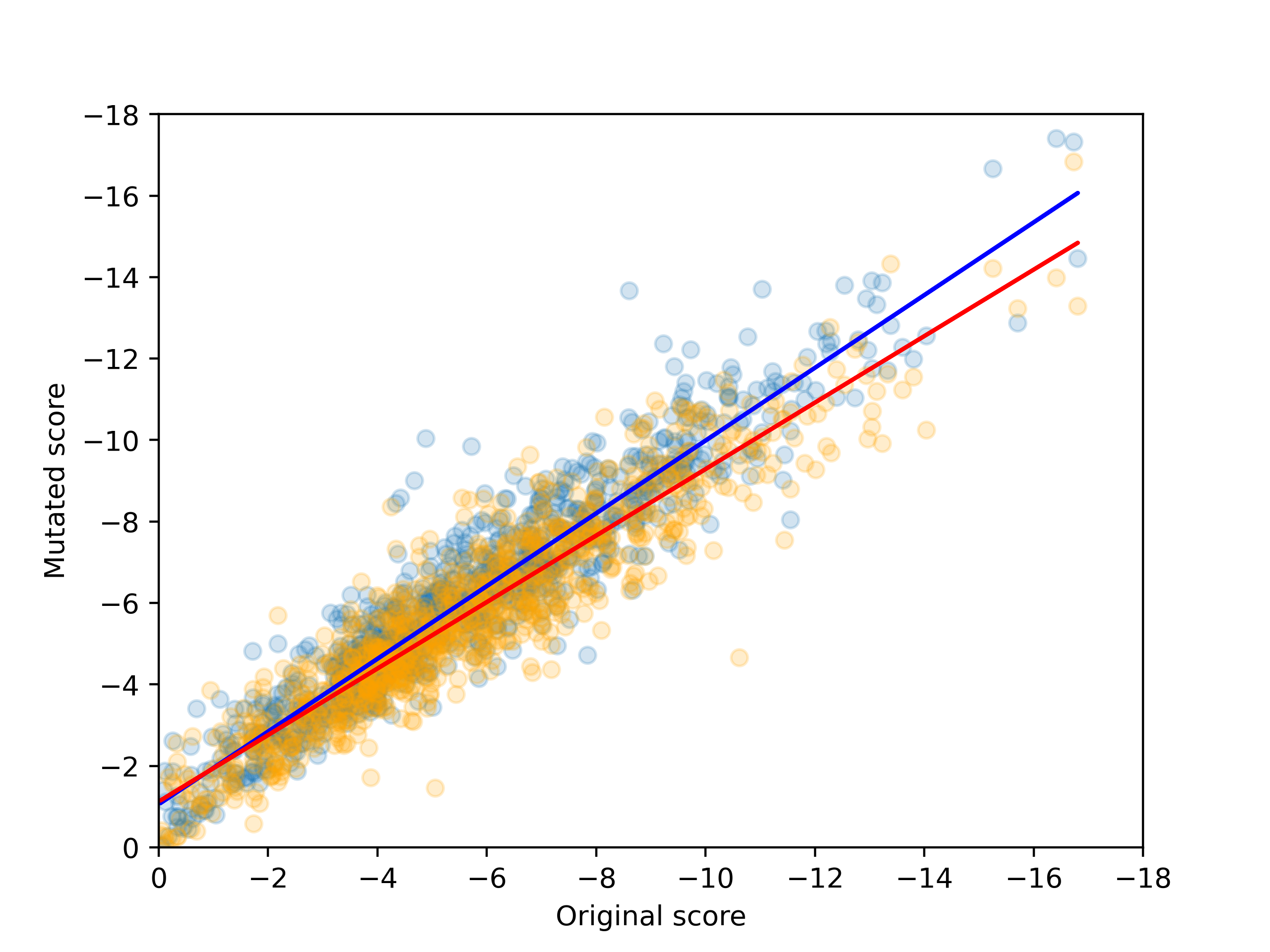}
\end{center}
\caption {X axis shows the free energies of piRNA (-RT ln Z) for the original sequences and Y axis shows the free energy when mutated. Blue shows the A/G mutation and orange shows A/T mutations. As you can see, the fitted line to the blue points is constantly above the red line (fitted to the orange points).  \label{fig:AG}} 
\end{figure}

\begin{center}
\begin{tabular}{ c | c | c | c }
 Mutation & C/G & C/T & C/A\\ 
 \hline
 Decrease of the free energy & -0.14 & -0.92 & -3.12\\ 
 
\end{tabular}
\end{center}

\begin{figure}[h!]
\begin{center}
\includegraphics[width=0.55\textwidth]{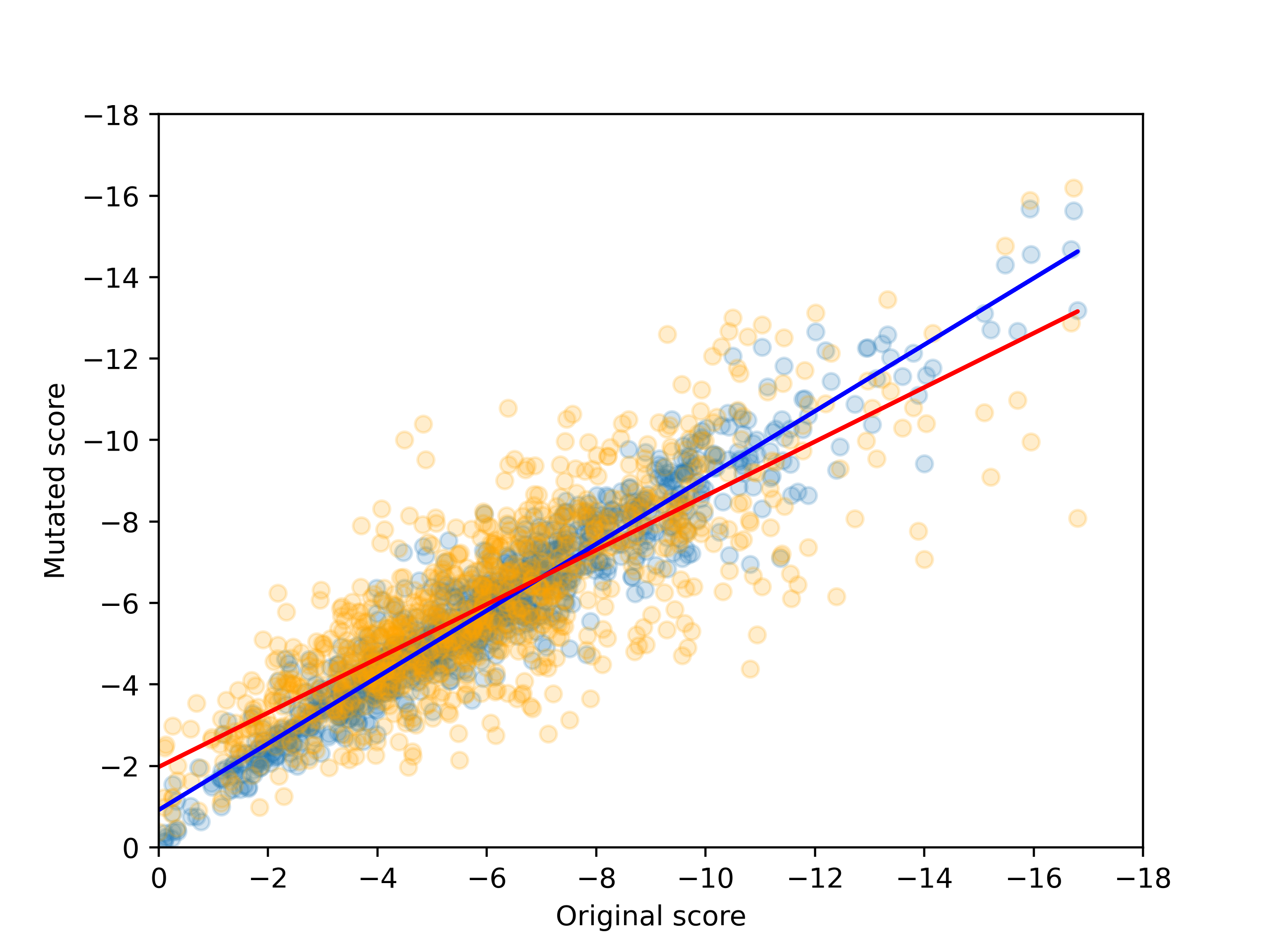}
\end{center}
\caption {X axis shows the free energies of piRNA (-RT ln Z) for the original sequences and Y axis shows the energy when mutated. Blue shows the C/T mutation and orange shows C/G mutations. As you can see for the more energetically favorable interaction sites (the ones toward the right side of the X axis which have lower free energy) the C/T mutation is still less detrimental on average and the fitted blue line is above the fitted red line after the intersection at around -8 kcal/mol. \label{fig:CT}} 
\end{figure}

\label{table_ref}

\begin{longtable}{ p{.15\textwidth}  p{.15\textwidth}  p{.15\textwidth}  p{.15\textwidth}  p{.15\textwidth}}
   & NCBI ID  & svRNA1 & svRNA2 & svRNA3 \\
   \hline
 1 & LC528232 & 3056 & 3182 & 28615 \\ 
 2 & LC528233 & 3061 & 3187 & 28620 \\ 
 3 & LC529905 & 3053 & 3179 & 28612 \\ 
 4 & LC529996 & 3033 & 3159 & 28592 \\ 
 5 & LR757995 & 3038 & 3164 & 28597 \\ 
 6 & LR757998 & 3028 & 3154 & 28587 \\ 
 7 & MN908947 & 3053 & 3179 & 28612 \\ 
 8 & MN938384 & 3021 & 3147 & 28580 \\ 
 9 & MN975262 & 3053 & 3179 & 28612 \\ 
 10 & MN985325 & 3053 & 3179 & 28612 \\ 
 11 & MN988668 & 3052 & 3178 & 28611 \\ 
 12 & MN988669 & 3052 & 3178 & 28611 \\ 
 13 & MN988713 & 3053 & 3179 & 28612 \\ 
 14 & MN994467 & 3053 & 3179 & 28612 \\ 
 15 & MN994468 & 3053 & 3179 & 28612 \\ 
 16 & MN996527 & 3020 & 3146 & 28579 \\ 
 17 & MN996528 & 3053 & 3179 & 28612 \\ 
 18 & MN996529 & 3041 & 3167 & 28600 \\ 
 19 & MN996530 & 3039 & 3165 & 28598 \\ 
 20 & MN996531 & 3040 & 3166 & 28599 \\ 
 21 & MN997409 & 3053 & 3179 & 28612 \\ 
 22 & MT007544 & 3053 & 3179 & 28612 \\ 
 23 & MT012098 & 3040 & 3166 & 28596 \\ 
 24 & MT019529 & 3053 & 3179 & 28612 \\ 
 25 & MT019530 & 3053 & 3179 & 28612 \\ 
 26 & MT019531 & 3053 & 3179 & 28612 \\ 
 27 & MT019532 & 3053 & 3179 & 28612 \\ 
 28 & MT019533 & 3053 & 3179 & 28612 \\ 
 29 & MT020781 & 3053 & 3179 & 28612 \\ 
 30 & MT020880 & 3053 & 3179 & 28612 \\ 
 31 & MT020881 & 3053 & 3179 & 28612 \\ 
 32 & MT027062 & 3053 & 3179 & 28612 \\ 
 33 & MT027063 & 3053 & 3179 & 28612 \\ 
 34 & MT027064 & 3053 & 3179 & 28612 \\ 
 35 & MT039873 & 3050 & 3176 & 28609 \\ 
 36 & MT039887 & 3053 & 3179 & 28609 \\ 
 37 & MT039888 & 3053 & 3179 & 28612 \\ 
 38 & MT039890 & 3053 & 3179 & 28612 \\ 
 39 & MT044257 & 3053 & 3179 & 28612 \\ 
 40 & MT044258 & 3029 & 3155 & 28588 \\ 
 41 & MT049951 & 3053 & 3179 & 28612 \\ 
 42 & MT050493 & 3033 & 3159 & 28592 \\ 
 43 & MT066156 & 3053 & 3179 & 28612 \\ 
 44 & MT066175 & 3053 & 3179 & 28612 \\ 
 45 & MT066176 & 3053 & 3179 & 28612 \\ 
 46 & MT072688 & 3038 & 3164 & 28597 \\ 
 47 & MT093571 & 3053 & 3179 & 28612 \\ 
 48 & MT093631 & 3040 & 3166 & 28599 \\ 
 49 & MT106052 & 3053 & 3179 & 28612 \\ 
 50 & MT106053 & 3053 & 3179 & 28612 \\ 
 51 & MT106054 & 3053 & 3179 & 28612 \\ 
 52 & MT118835 & 3053 & 3179 & 28612 \\ 
 53 & MT121215 & 3053 & 3179 & 28612 \\ 
 54 & MT123290 & 3056 & 3182 & 28615 \\ 
 55 & MT123291 & 3050 & 3176 & 28609 \\ 
 56 & MT123292 & 3053 & 3179 & 28612 \\ 
 57 & MT123293 & 3047 & 3173 & 28606 \\ 
 58 & MT126808 & 3053 & 3179 & 28612 \\ 
 59 & MT135041 & 3053 & 3179 & 28612 \\ 
 60 & MT135042 & 3053 & 3179 & 28612 \\ 
 61 & MT135043 & 3053 & 3179 & 28612 \\ 
 62 & MT135044 & 3053 & 3179 & 28612 \\ 
 63 & MT152824 & 3051 & 3177 & 28610 \\ 
 64 & MT159705 & 3053 & 3179 & 28612 \\ 
 65 & MT159706 & 3053 & 3179 & 28612 \\ 
 66 & MT159707 & 3053 & 3179 & 28612 \\ 
 67 & MT159708 & 3053 & 3179 & 28612 \\ 
 68 & MT159709 & 3053 & 3179 & 28612 \\ 
 69 & MT159710 & 3053 & 3179 & 28612 \\ 
 70 & MT159711 & 3053 & 3179 & 28612 \\ 
 71 & MT159712 & 3053 & 3179 & 28612 \\ 
 72 & MT159713 & 3053 & 3179 & 28612 \\ 
 73 & MT159714 & 3053 & 3179 & 28612 \\ 
 74 & MT159715 & 3053 & 3179 & 28612 \\ 
 75 & MT159716 & 3038 & 3164 & 28597 \\ 
 76 & MT159717 & 3053 & 3179 & 28612 \\ 
 77 & MT159718 & 3053 & 3179 & 28612 \\ 
 78 & MT159719 & 3053 & 3179 & 28612 \\ 
 79 & MT159720 & 3053 & 3179 & 28612 \\ 
 80 & MT159721 & 3053 & 3179 & 28612 \\ 
 81 & MT159722 & 3053 & 3179 & 28612 \\ 
 82 & MT163716 & 3053 & 3179 & 28612 \\ 
 83 & MT163717 & 3047 & 3173 & 28606 \\ 
 84 & MT163718 & 3053 & 3179 & 28612 \\ 
 85 & MT163719 & 3053 & 3179 & 28612 \\ 
 86 & MT184907 & 3053 & 3179 & 28612 \\ 
 87 & MT184908 & 3053 & 3179 & 28612 \\ 
 88 & MT184909 & 3053 & 3179 & 28612 \\ 
 89 & MT184910 & 3053 & 3179 & 28612 \\ 
 90 & MT184911 & 3053 & 3179 & 28612 \\ 
 91 & MT184912 & 3053 & 3179 & 28612 \\ 
 92 & MT184913 & 3053 & 3179 & 28612 \\ 
 93 & MT188339 & 2999 & 3125 & 28558 \\ 
 94 & MT188340 & 2999 & 3125 & 28558 \\ 
 95 & MT188341 & 2999 & 3125 & 28561 \\ 
 96 & MT192759 & 3026 & 3152 & 28585 \\ 
 97 & MT192765 & 3046 & 3172 & 28605 \\ 
 98 & MT192772 & 3053 & 3179 & 28612 \\ 
 99 & MT192773 & 3052 & 3178 & 28611 \\ 
 100 & MT226610 & 3053 & 3179 & 28612 \\ 
 101 & MT233519 & 2999 & 3125 & 28558 \\ 
 102 & MT233520 & 2999 & 3125 & 28558 \\ 
 103 & MT233521 & 2999 & 3125 & 28558 \\ 
 104 & MT233522 & 2999 & 3125 & 28558 \\ 
 105 & MT233523 & 2999 & 3125 & 28558 \\ 
 106 & MT240479 & 3017 & 3143 & 28576 \\ 
 107 & MT246449 & 2979 & 3105 & 28538 \\ 
 108 & MT246450 & 3022 & 3148 & 28581 \\ 
 109 & MT246451 & 3008 & 3134 & 28567 \\ 
 110 & MT246452 & 3046 & 3172 & 28605 \\ 
 111 & MT246453 & 2979 & 3105 & 28538 \\ 
 112 & MT246454 & 3044 & 3170 & 28603 \\ 
 113 & MT246455 & 3008 & 3134 & 28567 \\ 
 114 & MT246456 & 2996 & 3122 & 28555 \\ 
 115 & MT246457 & 2997 & 3123 & 28553 \\ 
 116 & MT246458 & 2927 & 3053 & 28486 \\ 
 117 & MT246459 & 3047 & 3173 & 28606 \\ 
 118 & MT246460 & 3053 & 3179 & 28612 \\ 
 119 & MT246461 & 3021 & 3147 & 28580 \\ 
 120 & MT246462 & 3053 & 3179 & 28612 \\ 
 121 & MT246463 & 2927 & 3053 & 28486 \\ 
 122 & MT246464 & 3002 & 3128 & 28561 \\ 
 123 & MT246465 & 2931 & 3057 & 28490 \\ 
 124 & MT246466 & 3047 & 3173 & 28606 \\ 
 125 & MT246467 & 3050 & 3176 & 28609 \\ 
 126 & MT246468 & 2993 & 3119 & 28552 \\ 
 127 & MT246469 & 3019 & 3145 & 28578 \\ 
 128 & MT246470 & 3035 & 3161 & 28594 \\ 
 129 & MT246471 & 3021 & 3147 & 28580 \\ 
 130 & MT246472 & 2968 & 3094 & 28527 \\ 
 131 & MT246473 & 2979 & 3105 & 28535 \\ 
 132 & MT246474 & 3043 & 3169 & 28602 \\ 
 133 & MT246475 & 3033 & 3159 & 28592 \\ 
 134 & MT246476 & 3014 & 3140 & 28573 \\ 
 135 & MT246477 & 3022 & 3148 & 28581 \\ 
 136 & MT246478 & 3040 & 3166 & 28599 \\ 
 137 & MT246479 & 2979 & 3105 & 28538 \\ 
 138 & MT246480 & 3052 & 3178 & 28611 \\ 
 139 & MT246481 & 3021 & 3147 & 28580 \\ 
 140 & MT246482 & 2996 & 3122 & 28555 \\ 
 141 & MT246483 & 2927 & 3053 & 28486 \\ 
 142 & MT246484 & 3021 & 3147 & 28580 \\ 
 143 & MT246485 & 2925 & 3051 & 28484 \\ 
 144 & MT246486 & 3021 & 3147 & 28580 \\ 
 145 & MT246487 & 3037 & 3163 & 28596 \\ 
 146 & MT246488 & 3026 & 3152 & 28585 \\ 
 147 & MT246489 & 3003 & 3129 & 28562 \\ 
 148 & MT246490 & 2997 & 3123 & 28556 \\ 
 149 & MT251972 & 3016 & 3142 & 28575 \\ 
 150 & MT251973 & 3049 & 3175 & 28608 \\ 
 151 & MT251974 & 3002 & 3128 & 28561 \\ 
 152 & MT251975 & 3025 & 3151 & 28584 \\ 
 153 & MT251976 & 3051 & 3177 & 28610 \\ 
 154 & MT251977 & 2979 & 3105 & 28538 \\ 
 155 & MT251978 & 3051 & 3177 & 28610 \\ 
 156 & MT251979 & 3002 & 3128 & 28561 \\ 
 157 & MT251980 & 3002 & 3128 & 28558 \\ 
 158 & MT253696 & 2999 & 3125 & 28558 \\ 
 159 & MT253697 & 2999 & 3125 & 28558 \\ 
 160 & MT253698 & 2999 & 3125 & 28558 \\ 
 161 & MT253699 & 2999 & 3125 & 28558 \\ 
 162 & MT253700 & 2999 & 3125 & 28558 \\ 
 163 & MT253701 & 2999 & 3125 & 28558 \\ 
 164 & MT253702 & 2999 & 3125 & 28558 \\ 
 165 & MT253703 & 2999 & 3125 & 28558 \\ 
 166 & MT253704 & 2999 & 3125 & 28558 \\ 
 167 & MT253705 & 2999 & 3125 & 28558 \\ 
 168 & MT253706 & 2999 & 3125 & 28558 \\ 
 169 & MT253707 & 2999 & 3125 & 28558 \\ 
 170 & MT253708 & 2999 & 3125 & 28558 \\ 
 171 & MT253709 & 2999 & 3125 & 28558 \\ 
 172 & MT253710 & 2999 & 3125 & 28558 \\ 
 173 & MT327745 & 3049 & 3175 & 28608 \\ 
 174 & MT328032 & 3053 & 3179 & 28612 \\ 
 175 & MT334563 & 3052 & 3178 & 28611 \\ 
 176 & MT345880 & 3006 & 3132 & 28565 \\ 
 177 & MT350251 & 3037 & 3163 & 28596 \\ 
 178 & MT350266 & 3053 & 3179 & 28612 \\ 
 179 & MT350282 & 3053 & 3179 & 28612 \\ 
 180 & MT359866 & 3043 & 3169 & 28602 \\ 
 181 & MT370944 & 3013 & 3139 & 28572 \\ 
 182 & MT370968 & 2989 & 3115 & 28548 \\ 
 183 & MT370975 & 2998 & 3124 & 28557 \\ 
 184 & MT371019 & 2998 & 3124 & 28557 \\ 
 185 & MT371024 & 2981 & 3107 & 28540 \\ 
 186 & MT371034 & 3006 & 3132 & 28565 \\ 
 187 & MT371035 & 2998 & 3124 & 28557 \\ 
 188 & MT371036 & 2998 & 3124 & 28557 \\ 
 189 & MT371037 & 2989 & 3115 & 28548 \\ 
 190 & MT371048 & 3053 & 3179 & 28612 \\ 
 191 & MT371568 & 2929 & 3055 & 28488 \\ 
 192 & MT371572 & 2945 & 3071 & 28504 \\ 
 193 & MT372481 & 3048 & 3174 & 28607 \\ 
 194 & MT374112 & 3051 & 3177 & 28610 \\ 
 195 & MT375470 & 3021 & 3147 & 28580 \\ 
 196 & MT385448 & 3050 & 3176 & 28609 \\ 
 197 & MT394529 & 3042 & 3168 & 28601 \\ 
 198 & MT394864 & 2999 & 3125 & 28558 \\ 
 199 & MT396242 & 3028 & 3154 & 28587 \\ 

\caption{ Start loci of the three svRNAs in various strains of SARS-CoV-2 genome.}\label{tab:match} 
\end{longtable}

\subsection{Potential targets}
We also analyzed the potential target genes of these svRNAs in human body. To this end, we considered more than 32,000 known RNAs that are transcribed in human body. We processed the sequence of each of these RNAs and removed the intron regions of each. Finally, we aligned the reverse complement of our svRNAs to all the exon sequences. A highly similar region to the reverse complement of an svRNA is highly complementary to the svRNA and hence suggests a high chance of interacting with that svRNA. Tables showing the results of that alignment for the regions that had an alignment score higher than $0.7$ with these three svRNAs are available in the Appendix Tables \ref{tab:targets1}, \ref{tab:targets2}, and \ref{tab:targets3}. To compute the score of the alignment, we used a reward of one (1) for a pair of matching nucleotides and a penalty of negative one (-1) for substitutions and insertion/deletions (indels). In the end, the total score was divided by the length of each svRNA to normalize the scores.


\section{Analysis}

Our proposed svRNAs for SARS-CoV-2 are highly conserved versions of the svRNAs of SARS-CoV. Presence of these svRNAs in all the 200 reference sequences that we used to test our hypothesis increases the possibility of correctness of our claim. Also, our proposed svRNAs occur at very similar loci in different reference sequences, and these loci are almost the same as the ones for the original svRNAs reported for SARS-CoV. The fact that the two viruses are in the same subgenus makes our hypothesis more plausible. However, still this hypothesis has to be verified experimentally.

As mentioned earlier, the complete tables showing the possible target RNAs of our proposed svRNAs are available in the appendix. However, it is worth mentioning some of them. The second highest match to the reverse complement of the first svRNA is HIF3A transcript which is a transcriptional regulator in adaptive response to low oxygen levels. Silencing this gene affects the reaction of the body in response to hypoxia. The second best match with the reverse complement of the second svRNA is MEX3B transcript which is a member of MEX3 translational regulators. MEX3 are RNA-binding proteins that are evolutionarily conserved and their \emph{in vivo} functions is yet to be fully characterized.

\section{Conclusion}
In this paper, we reported three potential svRNAs, which are orthologs of SARS-CoV svRNAs, in the SARS-CoV-2 genome. To validate our results, we confirmed the discovered orthologs are fully conserved in all the publicly available genomes of various strains of SARS-CoV-2; the loci at which the SARS-CoV-2 orthologs occur are close to the loci at which SARS-CoV svRNAs occur. Furthermore, our proposed svRNAs occur at very similar loci in different reference sequences, and these loci are almost the same as the ones for the original svRNAs reported in SARS-CoV.

We also reported potential targets for these svRNAs. We hypothesize that the discovered orthologs play a role in pathogenesis of SARS-CoV-2, and therefore, antagomir-mediated inhibition of these SARS-CoV-2 svRNAs inhibits COVID-19. This \emph{in silico} discovery still needs to be confirmed using \emph{in vitro} and \emph{in vivo} experiments.

\bibliographystyle{unsrt}
\footnotesize{\bibliography{masterref,pub}}

\newpage
\appendix
\section*{Appendix}

Tables~\ref{tab:targets1},~\ref{tab:targets2}, and~\ref{tab:targets3} show the matches with a score of at least 70\% between more than 32,000 RNAs transcribed in the human body and reverse-complement of our propossed svRNAs for Sars-CoV-2. The ones with a high score are likely to be target genes of the corresponding svRNA. The entries of each table are sorted based on the normalized score of the alignment.

Tables~\ref{tab:dnatargets1},~\ref{tab:dnatargets2}, and~\ref{tab:dnatargets3} show the matches with a score of at least 80\% between our proposed svRNAs and human reference genome. Patch release 13 of build 38 was used for as the reference genome.

\label{targets_1}

\begin{longtable}{ p{.03\textwidth}  p{.23\textwidth}  p{.05\textwidth}  p{.50\textwidth}  p{.05\textwidth}}
   & Gene ID  & Score &  Matched Sequence & Loci \\
\hline
 1 & ENSG00000168421 & 0.94 & ATCACCTTCTTCTTCACTC & 20 \\ 
\vspace{0.5cm} 
 & & & ATCACCTTCTTCTTCA-TC & \\ 
 2 & ENSG00000196218 & 0.89 & ATCACCTTCTTCTTCTTC & 15459 \\ 
\vspace{0.5cm} 
 & & & ATCACCTTCTTCTTCATC & \\ 
 3 & ENSG00000124440 & 0.89 & ATCACCTTCTTCTTCTTC & 5383 \\ 
\vspace{0.5cm} 
 & & & ATCACCTTCTTCTTCATC & \\ 
 4 & ENSG00000234545 & 0.83 & ATTCATCTTCTTCTTCATC & 1521 \\
\end{longtable}

\label{targets_2}

\begin{longtable}{ p{.02\textwidth}  p{.22\textwidth}  p{.03\textwidth}  p{.50\textwidth}  p{.05\textwidth}}
   & Gene ID  & Score &  Matched Sequence & Loci \\
   \hline
 1 & ENSG00000196628 & 0.86 & AGATCTTCTTCTTGCCTACTTCTTC & 10282 \\ 
\vspace{0.5cm} 
 & & & A-ATCTTCTTCTTG-CT-CTTCTTC & \\ 
 2 & ENSG00000183496 & 0.82 & TATCTTCTTCCTGCTCTTCTTC & 2447 \\ 
\vspace{0.5cm} 
 & & & AATCTTCTTCTTGCTCTTCTTC & \\ 
 3 & ENSG00000102290 & 0.77 & AAT-TTCTTCTTCCTCTTCTCTC & 1349 \\ 
\vspace{0.5cm} 
 & & & AATCTTCTTCTTGCTCTTCT-TC & \\ 
 4 & ENSG00000174175 & 0.77 & AATCATTCATTCTTG-TCTTCTTGC & 2614 \\
\end{longtable}

\label{targets_3}

\begin{longtable}{ p{.03\textwidth}  p{.23\textwidth}  p{.05\textwidth}  p{.50\textwidth}  p{.05\textwidth}}
   & Gene ID  & Score &  Matched Sequence & Loci \\
   \hline
 1 & ENSG00000166908 & 0.84 & ACGCTTCTGCCCCAGTTCCT & 759 \\ 
\vspace{0.5cm} 
 & & & A-GCTTCTGGCCCAGTTCCT & \\ 
 2 & ENSG00000174243 & 0.84 & AGCTTCTTGGCCCAGTT-CT & 4934 \\ 
\vspace{0.5cm} 
 & & & AGCTTC-TGGCCCAGTTCCT & \\ 
 3 & ENSG00000280011 & 0.84 & AGCATCTGGCCCCAGTTCCT & 7953 \\ 
\vspace{0.5cm} 
 & & & AGCTTCTGG-CCCAGTTCCT & \\ 
 4 & ENSG00000145685 & 0.84 & AG-TTCTGGCCCAGTCTCCT & 5046 \\
\end{longtable}

\label{dnatargets_1}

\begin{longtable}{ p{.03\textwidth}  p{.23\textwidth}  p{.05\textwidth}  p{.50\textwidth}  p{.05\textwidth}}
   & Gene ID  & Score &  Matched Sequence & Loci \\
\hline
 1 & NC000002.12 chr2 & 0.89 & GAGGAAGAAGAAGGTGAT & 134222364 \\ 
\vspace{0.5cm} 
 & & & GATGAAGAAGAAGGTGAT & \\ 
 2 & NC000003.12 chr3 & 0.89 & GATAAAGAAGAAGGTGAT & 62430918 \\ 
\vspace{0.5cm} 
 & & & GATGAAGAAGAAGGTGAT & \\ 
 3 & NC000012.12 chr12 & 0.89 & GATGCAAGATAGAAGGTGAT & 43685381 \\ 
\vspace{0.5cm} 
 & & & GATG-AAGA-AGAAGGTGAT & \\ 
 4 & NC000012.12 chr12 & 0.89 & GAT-AAGAAGAAGGTGAT & 22739819 \\
\end{longtable}

\label{dnatargets_2}

\begin{longtable}{ p{.02\textwidth}  p{.22\textwidth}  p{.03\textwidth}  p{.50\textwidth}  p{.05\textwidth}}
   & Gene ID  & Score &  Matched Sequence & Loci \\
   \hline
 1 & NC000004.12 chr4 & 0.86 & GAAGAAGAGGAAGAAGAAGAATT & 117558370 \\ 
\vspace{0.5cm} 
 & & & GAAGAAGAGCAAGAAGAAG-ATT & \\ 
 2 & NC000014.9 chr14 & 0.86 & GAAGAAGAGCAATGAAGAAAGACTT & 12719463 \\ 
\vspace{0.5cm} 
 & & & GAAGAAGAGCAA-GAAG-AAGA-TT & \\ 
 3 & NW018654725.1 chrY patch FIX & 0.86 & GAAGAAGAGGAAGAAGAAAGATT & 21110 \\ 
\vspace{0.5cm} 
 & & & GAAGAAGAGCAAGAAG-AAGATT & \\ 
 4 & NC000004.12 chr4 & 0.86 & GAAGAAGAAG-AAGAAGAAGATT & 130552436 \\
\end{longtable}

\label{dnatargets_3}

\begin{longtable}{ p{.03\textwidth}  p{.23\textwidth}  p{.05\textwidth}  p{.50\textwidth}  p{.05\textwidth}}
   & Gene ID  & Score &  Matched Sequence & Loci \\
   \hline
 1 & NC000002.12 chr2 & 0.89 & AGGACTACTGGGCCAGAAGCT & 48975650 \\ 
\vspace{0.5cm} 
 & & & AGGA--ACTGGGCCAGAAGCT & \\ 
 2 & NC000007.14 chr7 & 0.89 & AGGAACTGAGGCCAGGAAGCT & 14547625 \\ 
\vspace{0.5cm} 
 & & & AGGAACTG-GGCCA-GAAGCT & \\ 
 3 & NC000008.11 chr8 & 0.89 & AGGAACTGGGCCA-AAGCT & 10614535 \\ 
\vspace{0.5cm} 
 & & & AGGAACTGGGCCAGAAGCT & \\ 
 4 & NC000014.9 chr14 & 0.89 & AGGAACTGGGCCAGAAGCGCT & 1419369 \\
\end{longtable}

\end{document}